\documentstyle[amssymb]{article}

\input{tcilatex}

\begin{document}

\title{Equilibrium and nonequilibrium solitons in a lossy split-step system with
lumped amplification}
\author{Rodislav Driben and Boris A. Malomed \\
Department of Interdisciplinary Studies,\\
Faculty of Engineering, Tel Aviv University, Tel Aviv 69978,\\
Israel}
\maketitle

\begin{center}
{\bf ABSTRACT}
\end{center}

We propose a more realistic version of the recently introduced split-step
model (SSM), which consists of periodically alternating dispersive and
nonlinear segments, by adding uniformly distributed loss and {\it lumped}
gain to it. In the case when the loss is exactly balanced by gain, a family
of stable equilibrium solitons (ESs) is found. Unless the system's period $L$
is very small, saturation is observed in the dependence of the amplitude of
the established ES vs. that of the initial pulse. Stable nonequilibrium
solitons (NESs) are found in the case when the gain slightly exceeds (by up
to $\simeq 3\%$) the value necessary to balance the loss. The existence of
NESs is possible as the excessive energy pump is offset by permanent
radiation losses, which is confirmed by computation of the corresponding
Poynting vector. Unlike ESs that form a continuous family of solutions, NES
is an isolated solution, which disappears in the limit $L\rightarrow 0$,
i.e., it cannot be found in the overpumped nonlinear Schr\"{o}dinger
equation. Interactions between ESs turn out to be essentially the same as in
SSM without loss and gain, while interactions between NESs are different:
two NESs perturb each other by the radiation jets emanating from them, even
if they are separated by a large distance. Moving NESs survive collisions,
changing their velocities.

\newpage

\section{Introduction}

It is commonly known that the observation and use of solitons in various
systems, such as fiber-optic telecommunication links \cite{HK}, makes it
necessary to periodically compensate losses. If the system is long, which is
the case for telecommunication links, or if it has the shape of a closed
loop which is composed of different elements (examples are fiber lasers
generating the so-called stretched pulses \cite{Haus}, and a ferromagnetic
film combined with an amplifier, which is able to support periodic trains of
propagating magnetic solitons \cite{Slavin}), the loss-compensating gain is
usually provided for by short (effectively, point-like) amplifying elements
which are periodically placed ({\it lumped}) in the case of the long link.

If the system itself, except for the lumped amplifiers, is uniform, the
discreteness of the gain is not an important factor: in that case, it is
enough to demand that the gain averaged along the link is in balance with
the uniformly distributed loss. In particular, if the undamped uniform
nonlinear system supports solitons, it is easy to demonstrate that the
balance condition provides for the stable existence of solitons in the
presence of loss and amplification \cite{review}. On the other hand, a
number of strongly heterogeneous systems, based on periodic alternation of
very different elements, have been recently introduced, chiefly in the
context of nonlinear optics. These include the well-known dispersion
management (i.e., alternation of fiber segments with opposite signs of the
group-velocity dispersion \cite{DM}), one- or multidimensional {\it tandem
systems} concatenating linear and quadratically nonlinear waveguide sections 
\cite{tandem}, media composed of alternating layers with different values of
the Kerr coefficient \cite{layered}, {\it alternate waveguides}, in which
waveguiding and anti-waveguiding nonlinear segments are juxtaposed \cite
{Gisin}, dynamical lattices subject to {\it diffraction management} \cite
{Mark}, and some others.

A new system belonging to this class was recently proposed in the form of a 
{\it split-step model} (SSM) \cite{SSM}. In SSM, pieces of a dispersive
optical fiber with negligible nonlinearity (in fact, these are not
necessarily fibers; they may instead be short elements with a very strong
group-velocity dispersion, such as a fiber Bragg grating \cite
{Braggcompensator} or a photonic crystal \cite{photcrystal}) periodically
alternate with nonlinear fiber segments operating close to the
zero-dispersion point, in which the dispersion is negligible (these may be
replaced by short elements in which strong effective Kerr nonlinearity is
created by means of the cascading mechanism in quadratically nonlinear
optical crystals \cite{Wise}).

In the limit when lengths of both dispersive and nonlinear segments are much
smaller than the soliton's dispersion length, the SSM model is tantamount to
the commonly known split-step algorithm simulating the nonlinear
Schr\"{o}dinger (NLS) equation \cite{NM}. However, on the contrary to the
split-step algorithm, in SSM the separation between dispersive and nonlinear
segments is a physical feature, rather than a numerical trick, so that
lengths of the segments are not small, being {\em comparable to or larger
than} the dispersion length of a pulse (soliton) that may propagate in the
system.

In the above-mentioned works, it has been demonstrated, by means of direct
numerical simulations, that, in the absence of loss, all those strongly
inhomogeneous periodic systems support very robust quasi-soliton pulses,
although it may be quite difficult to prove their existence and stability in
a rigorous form. The introduction of losses and compensating gain poses a
new problem for solitons in these systems, the most fundamental issue being
whether the straightforward lumped amplification, with the average gain
compensating the loss, may support {\em stable} soliton-like pulses.

The objective of the present work is to look for stable solitons in what
seems to be the simplest and, possibly, most fundamental model of the type
described above: SSM with uniform losses and periodically placed point-like
amplifiers. Very recently, it has been demonstrated that, in a more general
case, when the average gain considerably exceeds the loss rate (by several
per cent), SSM solitons {\em cannot} be stable \cite{Alex}. It was found
that, nevertheless, efficient stabilization of solitons can be achieved by
means of synchronous intensity modulators (these time-domain control
elements are used in fiber-optic systems \cite{SIM}).

In the present work, we focus on a more special case, that, by itself, may
be less relevant to direct applications to telecommunications, but appears
to have a fundamental interest for the understanding of pulse dynamics. It
is a case when the gain exactly compensates the loss in SSM, or the relative
excess gain $\Delta G/G$ is small enough (in fact, it must be $\,\Delta
G/G\,\,_{\sim }^{<}\,0.03$), and no control elements (intensity modulators
or frequency-domain filters) are added. We find stable solitons of two
different types. First, in the case when the average gain exactly
compensates the loss ($\Delta G=0$), we find stable {\it equilibrium
solitons }(ESs). Usually, stationary solitary pulses in dissipative systems,
supported by the balance between loss and gain, are isolated solutions,
typical examples being an exact unstable solitary-pulse solution to the
cubic complex Ginzburg-Landau (GL) equation \cite{PS} and its stable
counterpart (also an exact solution) in a system of linearly coupled cubic
and linear GL equations \cite{Javid}. A difference of ESs found in the
present model is that they form a {\em continuous family}, which may be
parameterized by their amplitude. More accurately, the amplitude $\eta _{
{\rm fin}}$ of the soliton in the established state is nearly equal to the
initial amplitude $\eta _{{\rm in}}$, provided that $\eta _{{\rm in}}$ is
not too large. For larger values of $\eta _{{\rm in}}$, saturation in the
dependence $\eta _{{\rm fin}}\left( \eta _{{\rm in}}\right) $ is observed.

If small excess gain is present, being limited to the above-mentioned
interval $\Delta G/G\,\,_{\sim }^{<}\,0.03$, stable {\it nonequilibrium
solitons} (NESs) are found. Unlike the equilibrium ones, they are indeed
isolated solutions with a uniquely defined value of the amplitude. Their
existence in the presence of the excess gain is explained by the fact that
they permanently emit small-amplitude radiation waves. We check that the
size of the corresponding Poynting vector (energy flux) exactly matches the
rate of the energy pump into the soliton.

In this paper, we also demonstrate the existence of stable moving ESs and
NESs in the same model, and consider collisions between them, which turn out
to be inelastic in the latter case, although NESs survive collisions.
Interactions between immobile solitons are also considered, with a
conclusion that NESs strongly perturb each other, via the radiation jets
that they continuously emit.

In fact, ESs are quite simple pulses, which smoothly go over into the usual
nonlinear Schr\"{o}dinger (NLS) solitons as the system's period is
vanishing. On the other hand, NESs appear to be novel solitary pulses which
are supported by the stable balance between gain, dissipative loss, {\em and}
the emission of radiation. NESs have no counterparts in the usual uniform
NLS equation: we demonstrate that they disappear if the SSM period vanishes
(they disappear too if the period becomes too large). They also disappear if
the loss parameter is very large, and they only exist within the
above-mentioned narrow margin of values of the excess gain.

\section{The model}

The linear lossy dispersive segment of the system is described by the damped
linear Schr\"{o}dinger equation, 
\begin{equation}
iu_{z}+\frac{1}{2}u_{\tau \tau }=-\alpha _{D}u.  \label{dispersion}
\end{equation}
In an optical fiber, $u$, $z$ and $\tau $ are, respectively, the local
amplitude of the electromagnetic waves, propagation distance, and local time
($\tau \equiv t-z/V$, where $t$ and $V$ are the time proper and mean group
velocity of the wave packet), the dispersion coefficient is normalized to be 
$1$, and $\alpha _{D}>0$ is a loss constant. Obviously, Eq. (1) can be
solved by means of the Fourier transform in $\tau $ (which is used as a part
of the standard split-step numerical algorithm \cite{NM}).

The nonlinear segment of the system is governed by the dispersionless NLS\
equation, 
\begin{equation}
iu_{z}+|u|^{2}u=-i\alpha _{N}u,  \label{nonlinearity}
\end{equation}
where the nonlinearity coefficient is normalized to be $1$, and $\alpha
_{N}>0$ is the loss constant in the nonlinear segment (we assume that the
third-order dispersion may be neglected). A solution to Eq. (\ref
{nonlinearity}) is obvious, 
\begin{equation}
u(z,\tau )=u(0,\tau )\exp \left( -\alpha _{D}\,z\right) \exp \left[ i\frac{
|u(0,\tau )|^{2}}{2\alpha _{N}}\left[ 1-\exp \left( -2\alpha _{N}z\right) 
\right] \right] .  \label{phase}
\end{equation}

We define the system's elementary cell as an interval between midpoints of
two neighboring nonlinear segments. Point-like amplifiers are set at
junctions between cells. They acts on the wave field so that 
\begin{equation}
u(\tau )\mapsto u(\tau )\cdot e^{G}\,,  \label{G}
\end{equation}
where $G$ is the gain (the gain defined in terms of the signal's power and
measured in dB is $8.69G$). Thus, the full transformation ({\it map}) for
the pulse passing a cell can be represented as a superposition of two
nonlinear phase transformations (\ref{phase}) corresponding to the nonlinear
half-segments at the cell's edges, and the Fourier transform corresponding
to the dispersive segment in the middle of the cell, which are followed by
the multiplication as per Eq. (\ref{G}).

In order to prevent radiation from re-entering the integration domain, the
linear equation (\ref{dispersion}) was solved directly by means of numerical
methods in a broad integration interval $\Delta T$, placing absorbers at its
edges. In fact, the absorbers emulate a real physical effect, namely, the
fact that the energy emitted with radiation is lost.

If the lengths of the dispersive and nonlinear segments are $L_{D}$ and 
$L_{N}$, the cell size is $L=L_{D}+L_{N}$. Note that the left-hand sides of
Eqs. (\ref{dispersion}) and (\ref{nonlinearity}) are separately invariant
with respect to transformations 
\begin{equation}
\tau \rightarrow \tau /\Lambda _{D},z\rightarrow z/\Lambda _{D}^{2},\,{\rm
and}\,\,u\rightarrow \Lambda _{N}u,z\rightarrow z/\Lambda _{N}^{2}\,,
\label{scaling}
\end{equation}
The rescalings (\ref{scaling}) may be used (with $\Lambda _{D}^{2}/\Lambda
_{N}^{2}=L_{D}/L_{N}$) to make $L_{D}=L_{N}\equiv L/2\,$, which we assume
below to hold.

With regard to the normalizations adopted, an average ({\it distributed})
version of the model amounts to a perturbed NLS equation, 
\begin{equation}
iu_{z}+\frac{1}{4}u_{\tau \tau }+\frac{1}{2}|u|^{2}u=-i\left[ \frac{1}{2}
\left( \alpha _{N}+\alpha _{D}\right) -\frac{G}{L}\right] .  \label{NLS}
\end{equation}
The balance between the loss and gain is expected in the case when the
right-hand side of Eq. (\ref{NLS}) vanishes, i.e., when the gain takes the
equilibrium value 
\begin{equation}
G_{{\rm eq}}=\frac{L}{2}\left( \alpha _{N}+\alpha _{D}\right) .  \label{eq}
\end{equation}

\section{Equilibrium solitons}

As it was demonstrated in Ref. \cite{SSM}, SSM without loss and gain gives
rise to a family of stable solitary pulses. The first objective of the
present work is to generate similar pulses in the model introduced in the
previous section, in the case when the loss and gain are balanced as per Eq.
(\ref{eq}). To this end, we simulated the evolution of initial pulses that
were taken as fundamental solitons of the unperturbed NLS equation 
(\ref{NLS}), 
\begin{equation}
u(z=0,\tau )=\eta \,{\rm sech}\left( \eta \tau \right) ,  \label{initial}
\end{equation}
with an arbitrary amplitude $\eta $.

In a vast parametric range, the initial pulse (\ref{initial}) readily
generates a solitary wave, which remains stable as long as simulations could
be run, see a typical example in Fig. 1. As is seen in the figure, the pulse
changes its shape against the initial configuration (\ref{initial}).
Detailed analysis shows that the established pulse very well fits to the
expression (\ref{initial}), but with a different (smaller) value of the
amplitude. As well as the fundamental NLS soliton, the established pulse has
no {\it chirp} (i.e., no intrinsic phase structure \cite{HK}).

The family of {\it equilibrium solitons}, alias ESs [the ones subject to the
equilibrium condition \ref{eq})] found this way is characterized by a
dependence of the final (``output'') value $\eta _{{\rm out}}$ of the
amplitude on the initial (``input'') amplitude $\eta _{{\rm in}}$, which is
presented, for different fixed values of the period $L$, and for different
values of the loss constant, in Figs. 2(a) and 2(b). As is seen, for small
values of $\eta _{{\rm in}}$ the amplitude $\eta _{{\rm out}} $ is virtually
identical to it. However, with the increase of $\eta _{{\rm in}}$ the growth
of $\eta _{{\rm out}}$ slows down, and, eventually, $\eta _{{\rm out}}$
saturates at a constant value $\eta _{{\rm sat}}$, that strongly depends on 
$L$, but very weakly on $\alpha $, see Fig. 2(b). An example of the formation
of a stable ES with the amplitude $\eta _{{\rm out}}$ which is much smaller
than $\eta _{{\rm in}}$ (in the case of strong saturation) is displayed in
Fig. 3.

Figure 2(a) shows that, in the limit of very small $L$, the saturation does
not take place, and in this case we simply have $\eta _{{\rm out}}=\eta_{
{\rm in}}$\thinspace , which has a simple meaning: for small $L$, the
averaging applies, transforming the SSM with the balanced loss and gain into
the unperturbed NLS equation [see Eq. (\ref{NLS})]. In the latter equation,
the initial pulse (\ref{initial}) generates a fundamental soliton with the
amplitude $\eta $.

We stress that different branches of the dependence shown in Fig. 2(a) were
not aborted arbitrarily: except for the one corresponding to $L=0.1$, they
all terminate at points corresponding to a maximum value of $\eta _{{\rm in}
} $ for which the initial pulse (\ref{initial}) can generate a soliton. If 
$\eta _{{\rm in}}$ exceeds this maximum value, the initial pulse blows up,
generating spatio-temporal ``turbulence''. We did not study the latter
regime in detail, focusing on the regular behavior.

\section{Nonequilibrium solitons}

Numerical simulations of the model demonstrate that, if the gain slightly
exceeds the equilibrium value (\ref{eq}), stable solitons are still
generated in a narrow interval of values of the excess gain: $\left( G-G_{
{\rm eq}}\right) /G_{{\rm eq}}\lesssim 0.03$. A typical example of the
formation of such a {\it nonequilibrium} soliton (NES) is shown in Fig. 4,
which corresponds to $\left( G-G_{{\rm eq}}\right) /G_{{\rm eq}}=0.024$. If
the relative excess gain exceeds the maximum value $\simeq 0.03$, the system
blows up.

The apparent existence of the stable pulse shown in Fig. 4 suggests a
natural question: how is the energy balance provided for in this case, if
the gain exceeds the loss? An answer is that the energy pump into NES by the
excess gain is compensated by extra loss in the form of small-amplitude
radiation waves emitted by the pulse. Indeed, the background on top of which
the NES showed in Fig. 4 is found is not just $u\equiv 0$, but a
small-amplitude wave field.

To analyze this issue, we define the norm of the solution (it is what is
usually called energy in optics), $E=\int_{-\infty }^{+\infty }\left|
u(z,\tau )\right| ^{2}d\tau \,$. The energy is an obvious dynamical
invariant in the case of the unperturbed NLS equation, as well as in the SSM
without loss and gain. In the present model, the evolution of the energy of
a localized pulse is governed by a balance equation. After straightforward
calculations which use integration by parts, the balance equation can be
cast in the form 
\begin{equation}
\frac{dE_{{\rm sol}}}{dz}=2\left[ G\sum_{n}\delta (z-nL)-\alpha (z)\right]
E_{{\rm sol}}-D(z)\cdot {\rm Im}\left( u^{\ast }u_{\tau }\right) |_{\tau
=-T}^{\tau =+T}\,.  \label{balance}
\end{equation}
Here, the soliton's energy is defined as the integral over some finite
interval, $-T<\tau <+T$, which is essentially larger than a proper size of
the soliton, but smaller than the total size of the domain between the
above-mentioned edge absorbers, $E_{{\rm sol}}\equiv \int_{-T}^{+T}\left|
u(z,\tau )\right| ^{2}d\tau \,$. The last term in Eq. (\ref{balance}), which
is produced by the integration by parts, represents the {\it Poynting vector}
(energy flux) of the field $u$ at the points $\tau =\pm T$. Lastly, Eq. (\ref
{balance}) assumes that the amplifiers are placed at the points $z=Ln$, the
coefficient $\alpha (z)$ takes values $\alpha _{N}$ and $\alpha _{D}$ in the
nonlinear and dispersive segments, respectively, while $D(z)\equiv 1$ inside
the dispersive segments, and $\equiv 0$ inside the nonlinear ones.

In the case when the coefficients $\alpha $ and $G$ are small, Eq. (\ref
{balance}) can be averaged over a large number of cells, reducing to 
\begin{equation}
\frac{dE_{{\rm sol}}}{dz}=\left[ 2LG-\left( \alpha _{N}+\alpha _{D}\right)
\right] E_{{\rm sol}}-\frac{1}{2}\left( J_{+}+J_{-}\right) \,,
\label{Poynting}
\end{equation}
where $J_{\pm }$ are $z$-averaged values of ${\rm Im}\left( u^{\ast }u_{\tau
}\right) $ at the points $\tau =\pm T$. Taking, for instance, the NES
displayed in Fig. 4, and assuming that its shape is exactly described by Eq.
(\ref{initial}), we find $E_{{\rm sol}}=3.844$, hence the first term on the
right-hand side of Eq. (\ref{Poynting}), i.e., the net rate of the energy
pump into the NES, is $0.0165$. On the other hand, numerical data yield the
value $\left( J_{+}+J_{-}\right) /2\approx 0.0155$ of the second term (the
net energy flux across the points $\tau =\pm T$). The comparison shows
reasonable agreement between the energy pump and emission rates; the
remaining difference may be explained by a deviation of the exact shape of
the field from the expression (\ref{initial}).

A drastic difference of NES from ES is that NES is an isolated solution,
which is found with a single value of its amplitude, while ESs form a
continuous family. In other words, NES is a unique {\em attractor} in the
system.

It is instructive to summarize data showing the NES amplitude as a function
of the period $L$ and dissipative constant $\alpha $, see Fig. 5. An
important inference, suggested by Fig. 5(a) and confirmed by many more
simulations, is that the amplitude diverges (i.e., the NES blows up) as 
$L\rightarrow 0$. This result means that NES cannot exist in the uniform NLS
equation with unbalanced gain, which is quite obvious by itself. Thus, we
conclude that the existence of NES is a distinctive feature of SSM. Note
that NES does not exist (actually, it blows up) past the point where the
curve terminates in Fig. 5(b).

\section{Interactions and collisions between solitons}

Interactions between solitons is the next natural issue to consider. To this
end, we note that, although Eq. (\ref{nonlinearity}) and, hence, SSM as a
whole, are {\em not} Galilean invariant, moving pulses can be readily
generated from the zero-velocity ones considered above, ``pushing'' them by
means of a multiplier $\exp \left( -i\chi \tau \right) $ with an arbitrary
real constant $\chi $, added at the point $z=0$, which gives rise a
soliton's velocity proportional to $\chi $ \cite{SSM}. The result is
formation of stable moving solitons, both ESs and NESs, see an example of
moving NESs below in Fig. 7. Apparently, the amplitude of a moving NES is
somewhat smaller than that of its quiescent counterpart, but this issue was
not studied in detail.

Simulations demonstrate that both the interaction between quiescent ESs, as
well as collision between moving ones, are almost identical to those which
were studied in detail in SSM without loss and gain in Ref. \cite{SSM}.
Namely, collisions are quasi-elastic, and two zero-velocity ESs do not
interact, provided that the separation between them exceeds a certain
minimum value.

The situation is different in the case of NESs. First, even if the
separation between two zero-velocity NESs is large, they strongly perturb
each other -- not via direct interaction, but rather by the radiation
``jets'' emitted by each soliton, see a typical example in Fig. 6. Note that
each soliton has a smaller amplitude than in isolation. Second, moving NESs
survive collisions, but, as it is seen in the example displayed in Fig. 7,
the pulses reappear after the collision with different values of the
velocity [in fact, the slope of their trajectories in the $\left( \tau
,z\right) $ plane increases], and with a smaller amplitude. However, the
amplitude gradually relaxes to the unique value selected by NES moving at a
given velocity.

\section{Conclusion}

We have added loss and gain to a model of a long periodic system consisting
of separated dispersive and nonlinear segments. It was shown that, if the
loss is exactly balanced by gain, a family of stable equilibrium solitons
exists. Unless the system's period is very small, saturation is observed in
the dependence of the amplitude of the established soliton vs. that of the
initial pulse. Stable nonequilibrium solitons were found in the case when
the gain slightly exceeds (by up to $\simeq 3\%$) the value necessary to
balance the loss. The existence of NESs is possible as the excessive energy
pump is offset by permanent radiation losses, which is confirmed by
computation of the corresponding Poynting vector. Unlike ESs that exist as a
continuous family of solutions, NES is an isolated one, and it disappears
when the system's period vanishes. Interactions between ESs were found to be
essentially the same as in the model without loss and gain, while
interactions between NESs are different: two NESs perturb each other by
their radiation jets, even if they are separated by a large distance. Moving
NESs survive collisions, with a change of the velocity.

\section*{Acknowledgements}

We appreciate useful discussions with M.J. Ablowitz and P.L. Chu.

\newpage

\section{Figure captions}

Fig. 1. An example of the establishment of a stable pulse of the equilibrium
type in SSM with the balanced loss and gain [$\alpha _{N}=\alpha _{D}=0.035$,
$L=1$, and, in accord with Eq. (\ref{eq}), $G=0.035$]. The initial pulse
is taken as the NLS soliton (\ref{initial}) with $\eta =1$. In this and
subsequent figures, $|u|^{2}$ (``power'') is shown vs. $\tau $ at the end of
each cell.

Fig. 2. The established (``output'') amplitude of the equilibrium soliton
vs. the initial (``input'') amplitude. In (a), $\alpha _{N}=\alpha _{D}$ is
fixed to be $0.035$; in (b), $L$ is fixed to be $1$, and $\alpha _{N}=\alpha
_{D}\equiv \alpha $.

Fig. 3. Formation of a stable equilibrium soliton in the case when Fig. 2
shows strong saturation. The initial amplitude in Eq. (\ref{initial}) is 
$\eta =4$, $L=1$, $\alpha _{N}=\alpha _{D}=0.035$.

Fig. 4. An example of the formation of a nonequilibrium soliton. The
amplitude of the input pulse and the SSM's parameters are the same as in the
case shown in Fig. 1, except for $G=0.0\,\allowbreak 358\,5$, which
corresponds to the relative excess gain $\left( G-G_{{\rm eq}}\right) /G_{
{\rm eq}}\approx 0.024$.

Fig. 5. The amplitude of the stable nonequilibrium soliton vs. the system's
period (``stepsize'') $L$ (a) and the loss parameter $\alpha _{N}=\alpha
_{D}\equiv \alpha $ (b). In both panels, the relative excess gain is fixed
to be $0.024$; in (a), $\alpha =0.035$, and in (b), $L=1$.

Fig. 6. Interaction of two nonequilibrium solitons (the same ones as in Fig.
4). The initial separation between them is $\Delta \tau =18$, the
full-width-at-half-maximum \cite{HK} of each one being initially $T_{{\rm
FWHM}}=2$. (a) The distribution of $\left| u(z,\tau )\right| ^{2}$; (b) the
same shown by means of level contours.

Fig. 7. An example of the collision between two nonequilibrium solitons.
Each moving soliton is generated from the quiescent one shown in Fig. 4,
multiplying it by $\exp \left( \pm i\tau /2\right) $. The side view of the
distribution of $\left| u(z,\tau )\right| ^{2}$ is displayed [in this case,
a full three-dimensional picture like those in Figs. 1, 3, 4, and 6(a) would
seem messy], the inset showing trajectories of centers of the two solitons
on the plane $\left( \tau ,z\right) $.

\end{document}